\documentclass{appolb}
\usepackage{graphicx}
\usepackage{hyperref}

\usepackage{amsmath}
\usepackage{cite}

\begin{document}
\title{Diffractive  di-hadron production at NLO within the shockwave formalism 
\thanks{ Presented at Diffraction and Low-x 2022}%
}
\author{Michael Fucilla, 
\address{Dipartimento di Fisica, Università della Calabria, I-87036 Arcavacata di Rende, Cosenza, Italy \\ 
Istituto Nazionale di Fisica Nucleare, Gruppo collegato di Cosenza,-87036 Arcavacata di Rende,
Cosenza, Italy and \\ 
Université Paris-Saclay, CNRS/IN2P3, IJCLab, 91405, Orsay, France }
\\[3mm]
{Andrey V. Grabovsky 
\address{Budker Institute of Nuclear Physics, 11, Lavrenteva avenue, 630090, Novosibirsk, Russia and \\ 
Novosibirsk State University, 630090, 2, Pirogova street, Novosibirsk, Russia }}
\\[3mm]
{ Lech Szymanowski
\address{National Centre for Nuclear Research (NCBJ), Pasteura 7, 02-093 Warsaw, Poland}}
\\[3mm]
{Emilie Li, Samuel Wallon
\address{Université Paris-Saclay, CNRS/IN2P3, IJCLab, 91405, Orsay, France}}
}

\maketitle
\begin{abstract}
We compute the next-leading-order cross-sections for  diffractive electro- or photoproduction of a pair of hadrons with large $p_T$, out of a nucleus or a nucleon. A hybrid factorization is used, mixing collinear and small-$x$ factorizations, more precisely shockwave formalism. We demonstrate the cancellation of divergences and extract the finite parts of the differential cross-section in general kinematics. 
\end{abstract}
  
\section{Introduction}

To uncover accurately gluon saturation  in nucleons and nuclei, precision observables of experimentally-relevant processes are essential. In the last few years, several processes have been investigated, in diffractive DIS, such as exclusive dijet production\cite{Boussarie:2014lxa, Boussarie:2016ogo,Boussarie:2019ero}, exclusive meson production \cite{Boussarie:2016bkq}, as well as, in inclusive DIS, the production of single hadron \cite{Bergabo:2022zhe}, double hadron \cite{PhysRevD.106.054035}  and dijet  \cite{Caucal:2021ent}. 
We propose here the  diffractive di-hadron production in $\gamma^{(*)} + p/A$ as another path to saturation. The results are built upon \cite{Boussarie:2016ogo} where the Next-Leading-Order (NLO) impact factors are computed in the shockwave formalism. We will emphasize on the cancellation of infrared (IR) divergences between the virtual, real, and counterterms contributions. 

\section{Theoretical framework }

We consider the inclusive production of a pair of hadrons with $\vec{p}_{h_1}^{\, 2} \sim \vec{p}_{h_2}^{\, 2} $ 
\begin{equation}
    \gamma^{(*)}(p_\gamma) + P(p_0) \rightarrow h_1(p_{h1}) + h_2 (p_{h2}) + X + P'(p_0') 
\end{equation} 
where $X$ stands for the other undetected particles on the projectile side. 
A hybrid factorization (shockwave, collinear) is used. 

The shockwave framework describes the interaction of the probe with the target, including saturation effects. The space-time dimension is $D = 2 + d = 4 + 2 \epsilon$.
We introduce two light-cone vectors $n_1, n_2$ that define the $+/-$ directions respectively and work in the $n_2 \cdot A = A^+ = 0$ gauge. The gluon field is decomposed into external (internal) field $b^\mu$ ($\mathcal{A}^\mu$), depending on the value of their $+$ momentum being below (above) an arbitrary cut-off $e^\eta p_\gamma^+$. We boost from the target rest frame to our working frame where the photon and target move ultra-relativistically and $p_0^- \sim p_\gamma^+ \sim \sqrt{s}$ with $s$ the center of mass frame of the photon and the target. The $b^\mu$ field then has the form $b^\mu(z) = b^-(\vec{x})\delta(x^+)n_2^\mu $. Wilson lines 
\begin{equation}
    U_{\vec{z}} = \mathcal{P} \exp \left\{ i g \int_{- \infty}^{+ \infty} d z^+ b^-(z)\right\}
\end{equation}
resum all order eikonal interactions with those fields.

All momenta in the projectile side are decomposed as 
\begin{equation}
p_i^\mu = x_i p_\gamma^+ n_1^\mu + \frac{p_i^2 + \vec{p}^{\, 2 }}{2 x_i p_\gamma^+} n_2^\mu + p_{i \perp}^\mu  \; . 
\end{equation} 

The Pomeron exchange between the probe and the target is represented by color-singlet operators built on Wilson lines, e.g. the dipole operator
\begin{equation}
    \mathcal{U}_{ij} =  \Tr\left(U_{\vec{z}_i}U_{\vec{z}_j}^\dag \right) - N_c\,.
\end{equation}
Those operators evolve according to the B-JIMWLK equation \cite{Balitsky:2001re, BALITSKY199699, Balitsky:1998kc,Balitsky:1998ya, Jalilian-Marian:1997qno,Jalilian-Marian:1997jhx,Jalilian-Marian:1997ubg,Jalilian-Marian:1998tzv,Kovner:2000pt,Weigert:2000gi, Iancu:2000hn, Ferreiro:2001qy, Iancu:2001ad}. 

Amplitudes are factorized between the impact factors and the non-perturbative matrix elements of those operators between the target in and out state. Collinear factorization describes the fragmentation part, thanks to the hard scale $\vec{p}_{h}^{\, 2} \gg \Lambda_{QCD}^2$. We also impose $\vec{p}^{\, 2} \gg \vec{p}_{h}^{\, 2}$, $\vec{p}$ being the relative transverse momentum of the two hadrons. This implies that they have a large separation angle, eliminating the possibility of them being produced from one single parton. 
From this theorem and the collinearity of the fragmenting parton and the produced hadrons, the LO cross-section is the convolution of Fragmentation Functions (FF) and coefficient functions 
\begin{eqnarray}  
&&\frac{d \sigma_{0JI}^{h_1 h_2}}{d x_{h_1} d x_{h_2}d^d \vec{p}_{h_1}d^d \vec{p}_{h_2}} = \sum_{q} \int_{x_{h_1}}^1 \frac{d x_q}{x_q} \int_{x_{h_2}}^1 \frac{d x_{\bar{q}}}{x_{\bar{q}}} \left(\frac{x_q}{x_{h_1}}\right)^d \left(\frac{x_{\bar{q}}}{x_{h_2}}\right)^d \nonumber \\
&& \times D_q^{h_1}\left(\frac{x_{h_1}}{x_q}\right) D_{\bar{q}}^{h_2}\left(\frac{x_{h_2}}{x_{\bar{q}}}\right) \frac{d\hat{\sigma}_{JI}}{d x_q d x_{\bar{q}} d^d \vec{p}_q d^d \vec{p}_{\bar{q}}} + (h_1 \leftrightarrow h_2) 
\label{LO cross_section}
\end{eqnarray}
expressed in terms of the partonic cross-section,  $J,I$ representing the photon polarization for the complex amplitude and the amplitude respectively. 

\section{NLO computations in a nutshell}

The NLO density matrix contains  all types of contributions depending on the nature of the impact factors, i.e 
\vspace{.2cm}
\begin{figure}[h!]
\begin{picture}(430,390)
\put(20,330){\includegraphics[scale=0.2]{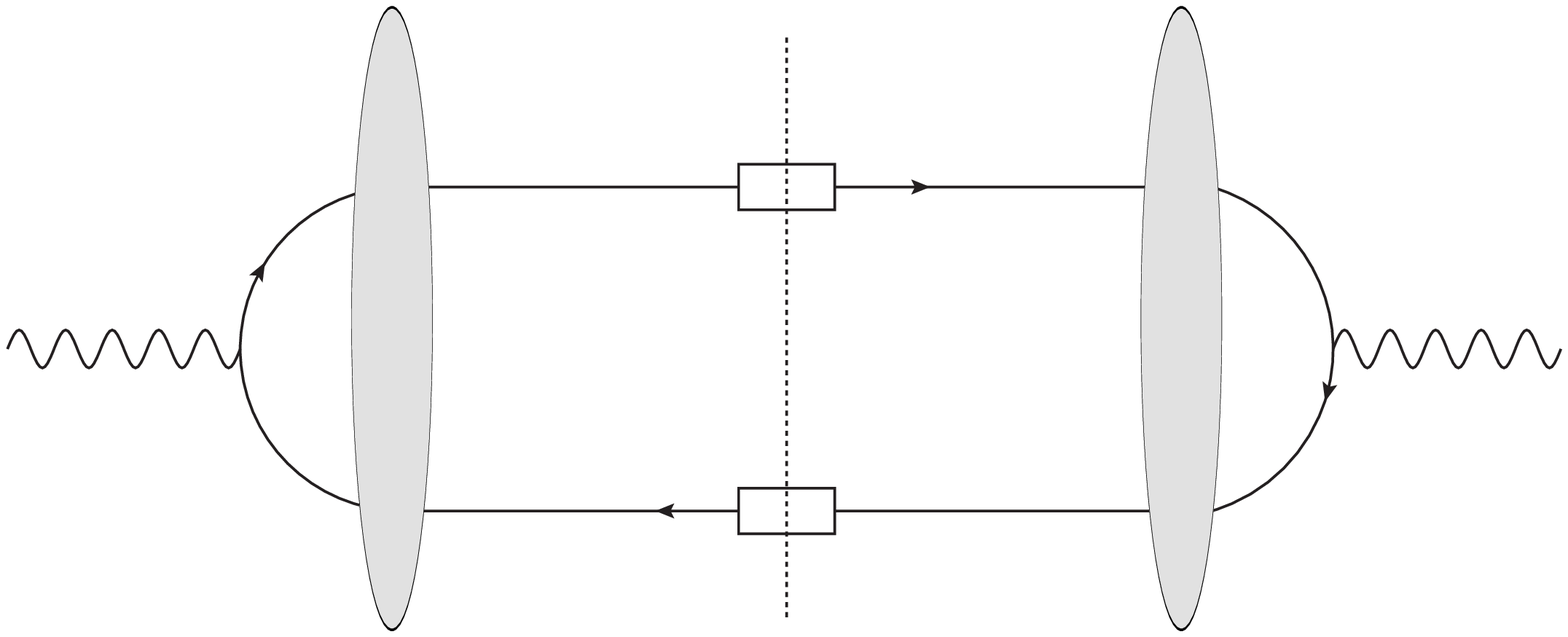}}
\put(73,395){ \tiny NLO}
\put(160,350){$=$}
\put(183,330){\includegraphics[scale=0.2]{images/FF_dihadron_LO_box.eps}}
\put(205,395){\tiny 1-loop}
\put(310,352){ \tiny + c.c}
\put(241,320){\small (a)}
\put(20,260){\includegraphics[scale=0.2]{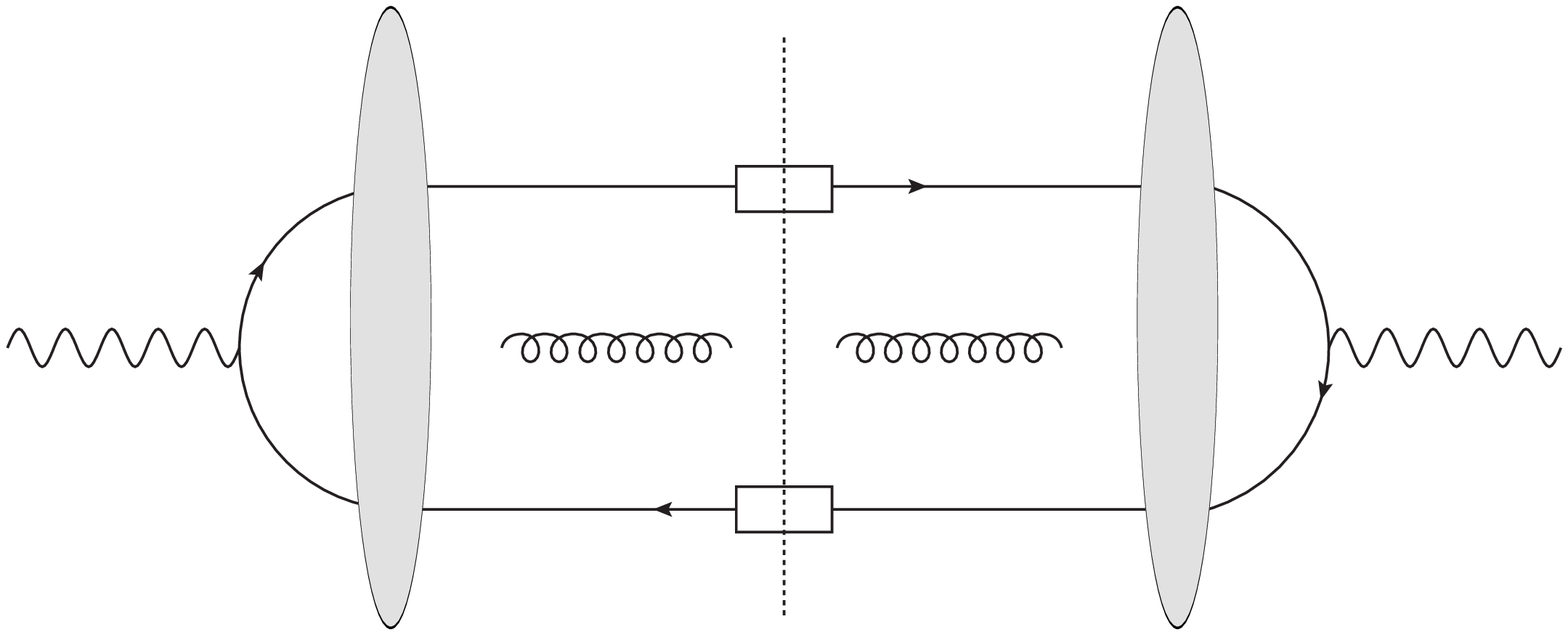}}
\put(8,282){ \tiny +}
\put(78,250){\small (b)}
\put(183,260){\includegraphics[scale=0.2]{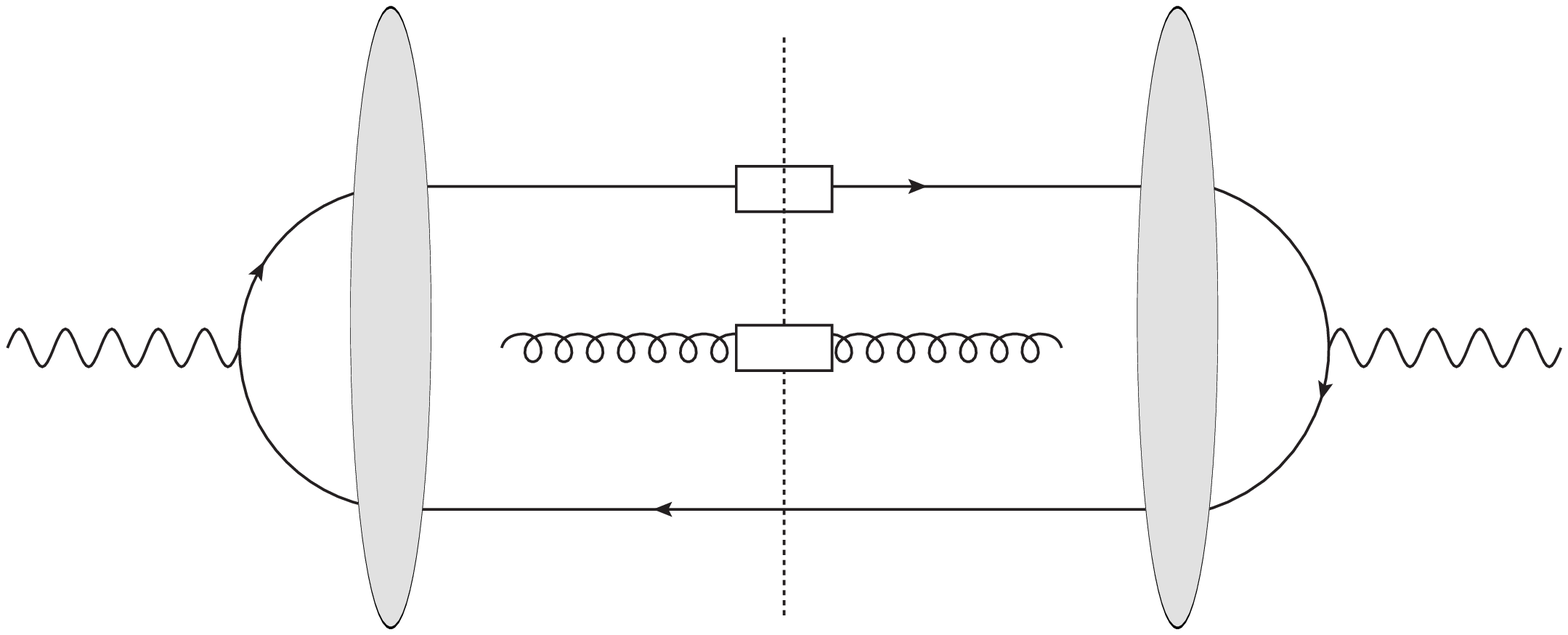}}
\put(162,282){\tiny +}
\put(20,190){\includegraphics[scale=0.2]{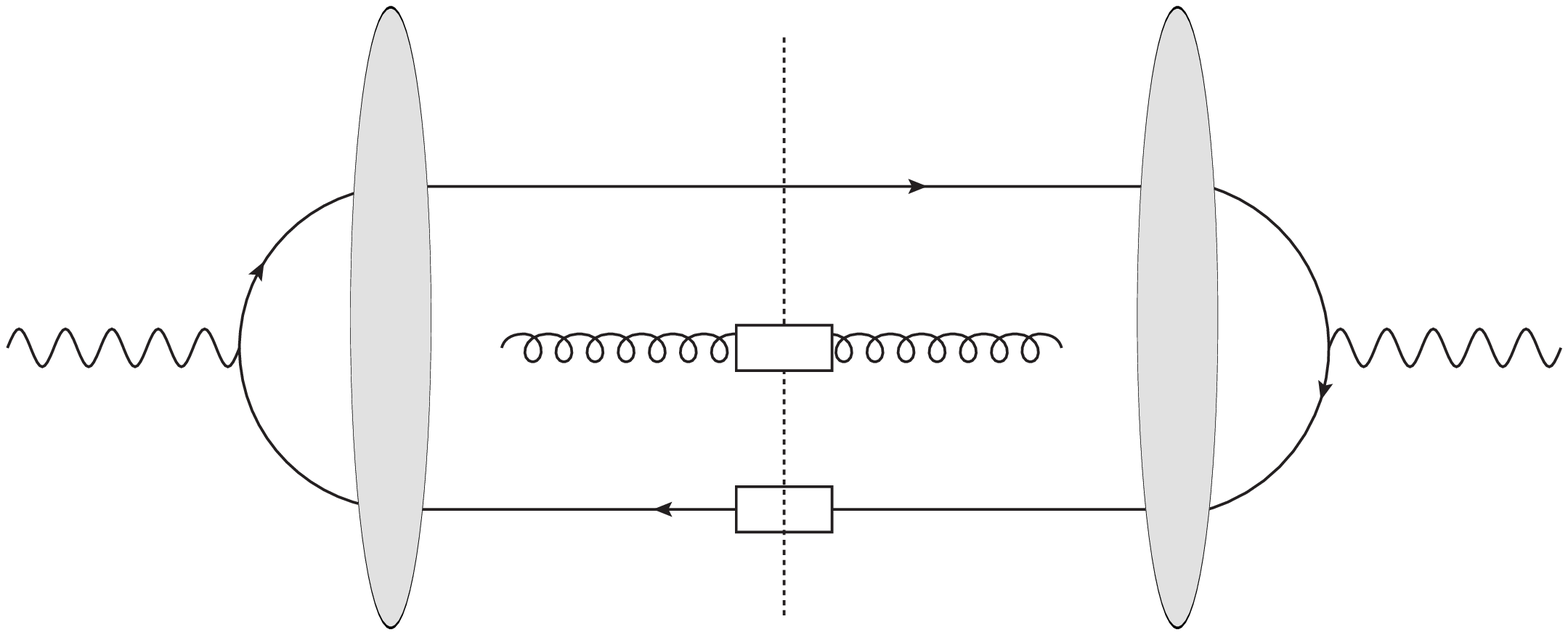}}
\put(241,250){\small (c)}
\put(75,180){ \small(d)}
\put(12, 212){\tiny +}
\put(162, 212){\tiny +}
\put(183,190){\includegraphics[scale=0.2]{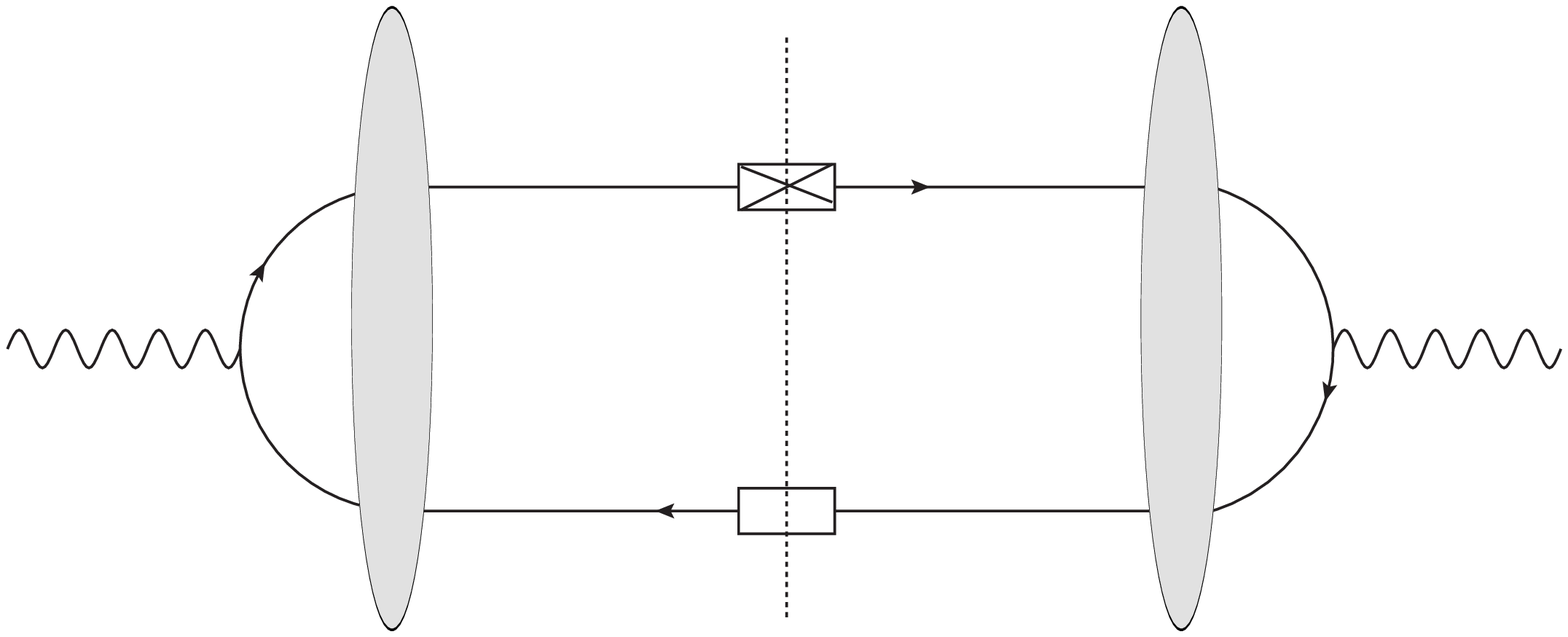}}
\put(241,180){(e)}
\put(178, 210){\Bigg ( }
\put(308, 212){ \tiny $+ \, q \leftrightarrow \bar{q}$}
\put(333,210){ \Bigg )}
\end{picture}
\vspace{-185pt}
  \caption{NLO cross-section dependence on FF, represented by the box.}
  \label{fig:NLO_FF}
\end{figure}
\begin{equation}
    d \sigma_{JI}^{NLO} = d \sigma_{1 JI} +  d \sigma_{2 JI}  +  d \sigma_{3 JI}  +  d \sigma_{4 JI}  +  d \sigma_{5 JI}  \; . 
\end{equation}
Here $d \sigma_{1 JI}$ and $d\sigma_{2 JI}$ are the dipole $\times$ dipole and dipole $\times$ double dipole virtual contributions while $d\sigma_{3 JI}$, $d\sigma_{4 JI}$  and $d\sigma_{5 JI}$ are the dipole $\times$ dipole, dipole $\times$  double dipole and double dipole $\times$ double dipole real parts. The various contributions depend on the detail of the FF used, as shown by fig~\ref{fig:NLO_FF}.

\begin{figure}[h!]
\begin{picture}(430,385)
\put(20,330){\includegraphics[scale=0.2]{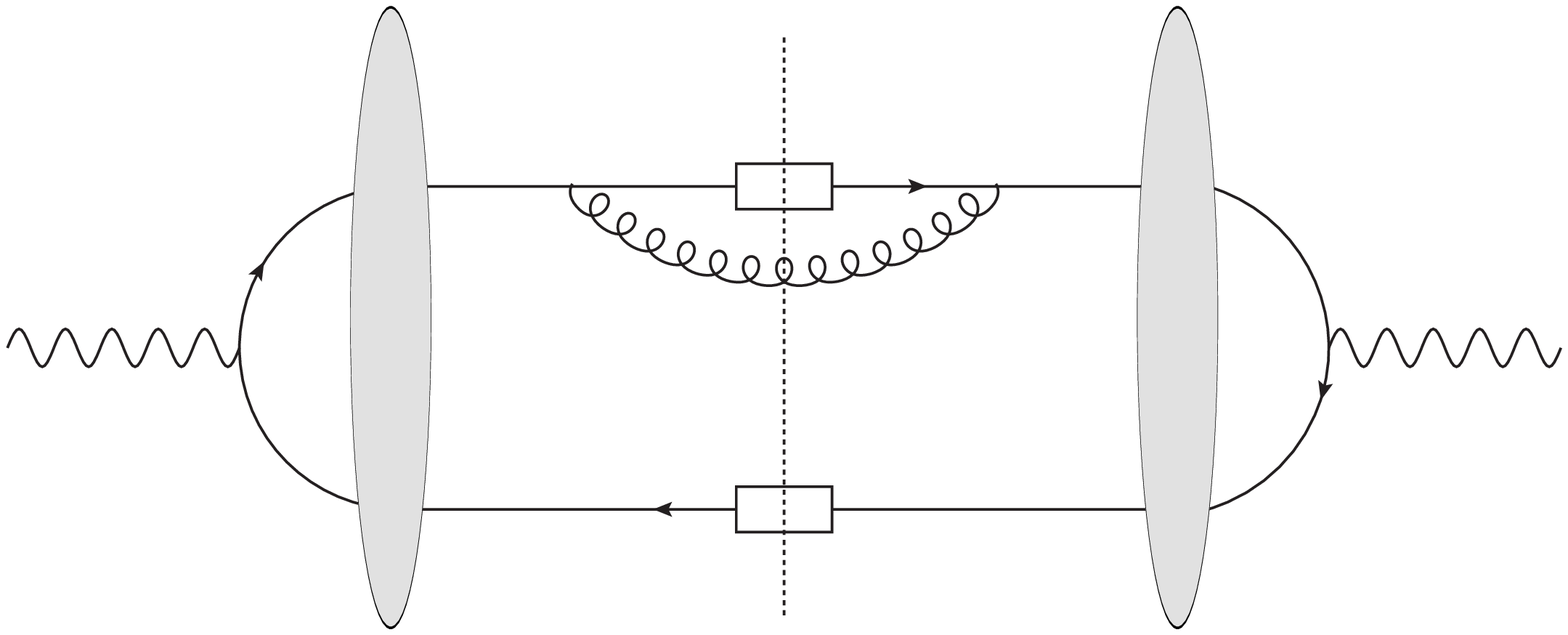}}
\put(37,320){\small (1) : soft + collinear ($qg$) }
\put(183,330){\includegraphics[scale=0.20]{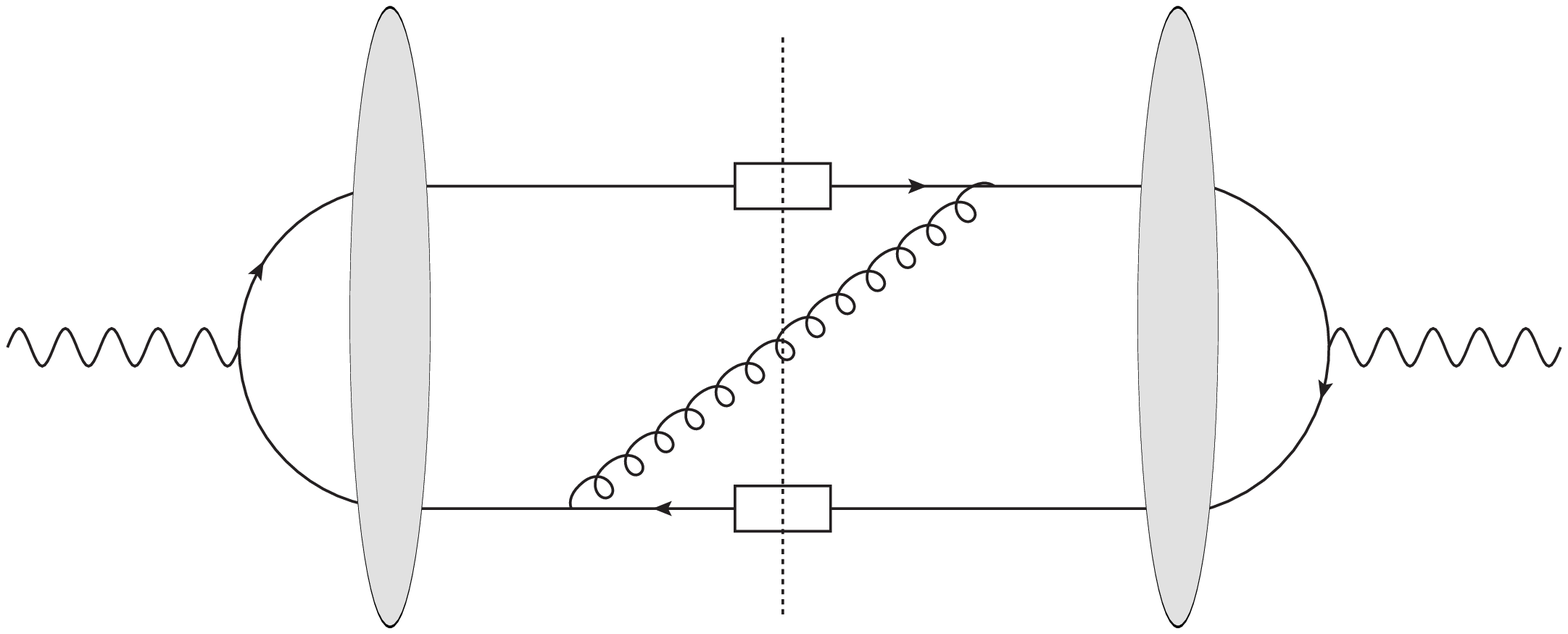}}
\put(230,320){\small (2) : soft }
\put(20,260){\includegraphics[scale=0.20]{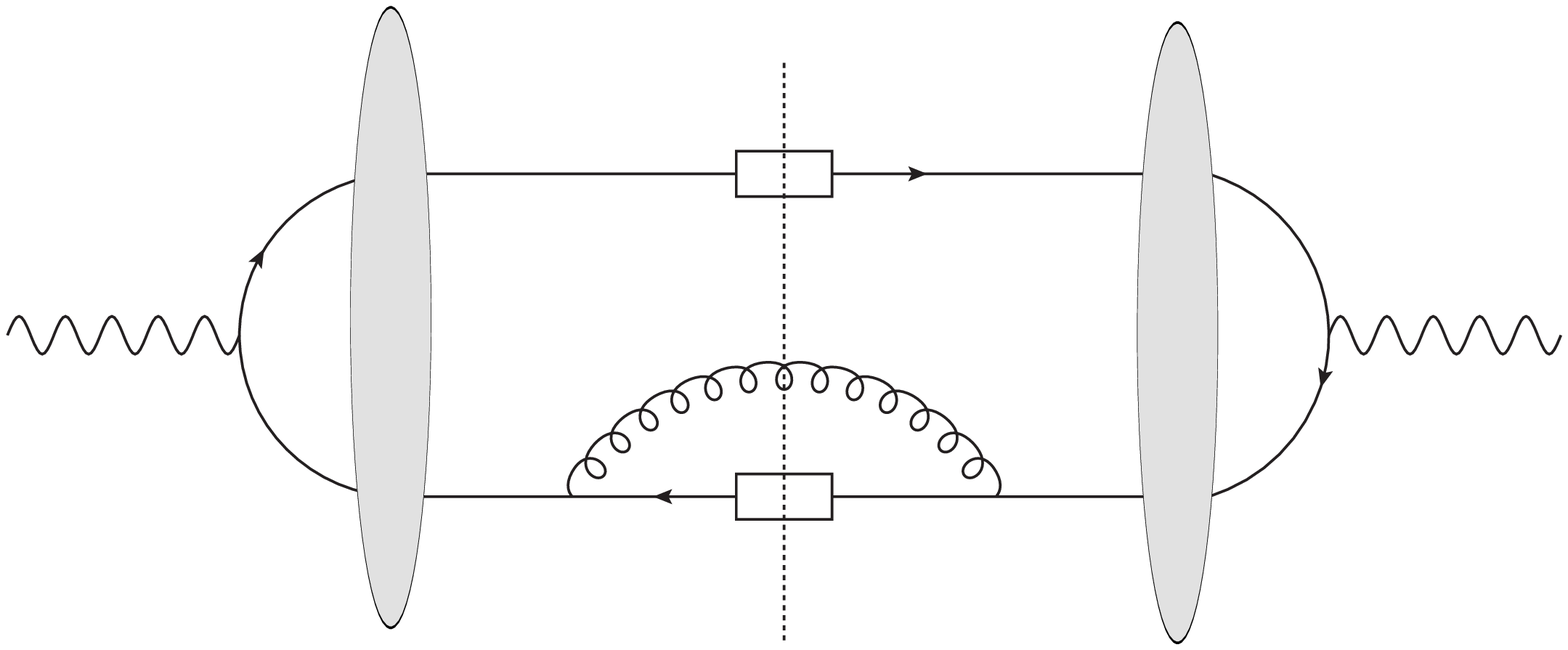}}
\put(37,250){\small (3) : soft + collinear ($\bar{q}g$)  }
\put(183,260){\includegraphics[scale=0.20]{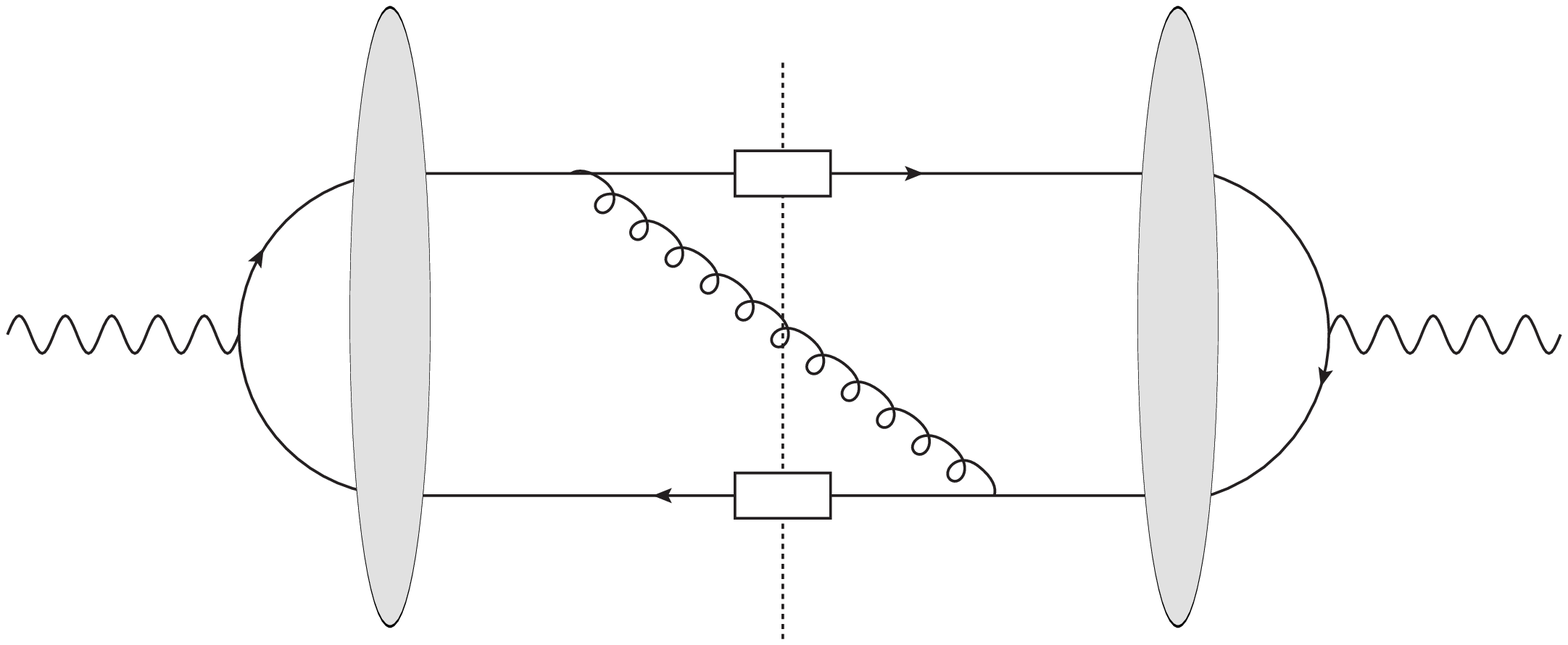}}
\put(230,250){\small (4) : soft }
\put(20, 190){\includegraphics[scale=0.20]{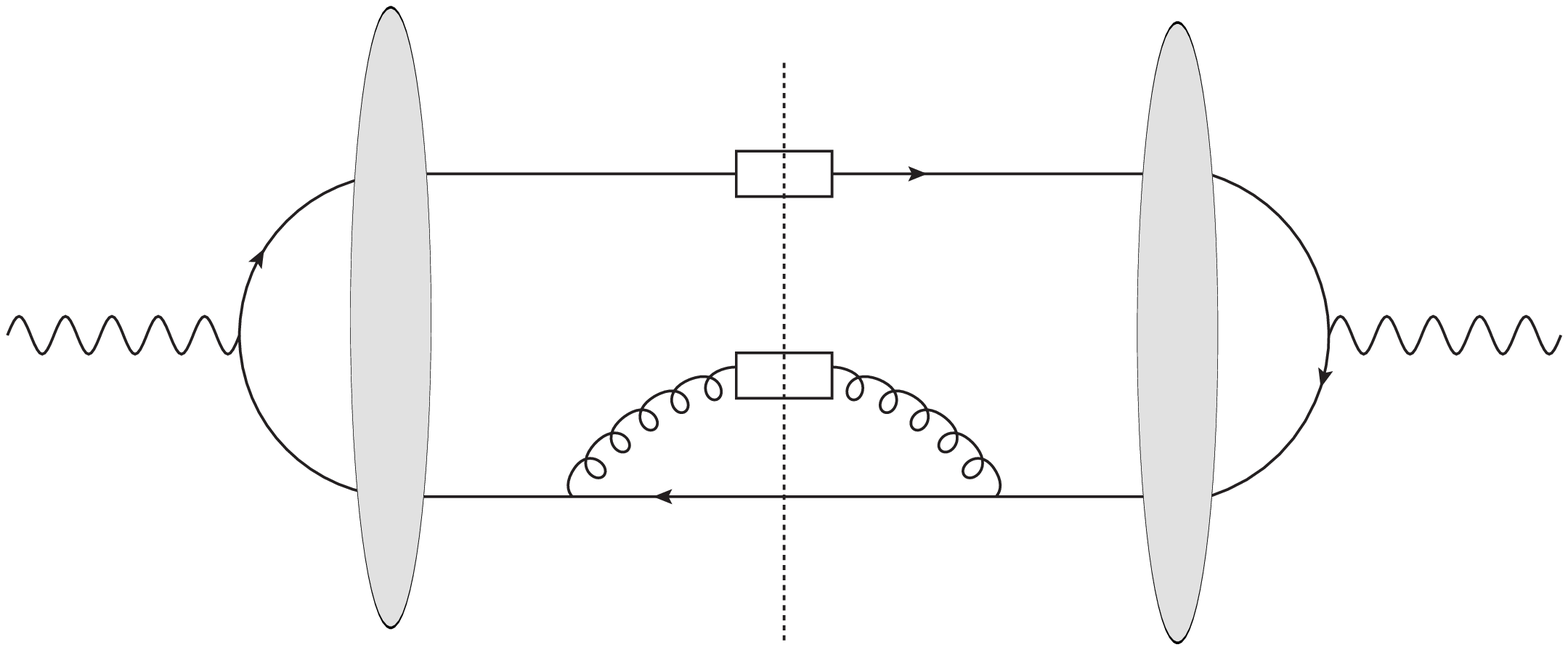}}
\put(43,180){\small (5) : collinear ($\bar{q}g$)  }
\put(183,190){\includegraphics[scale=0.20]{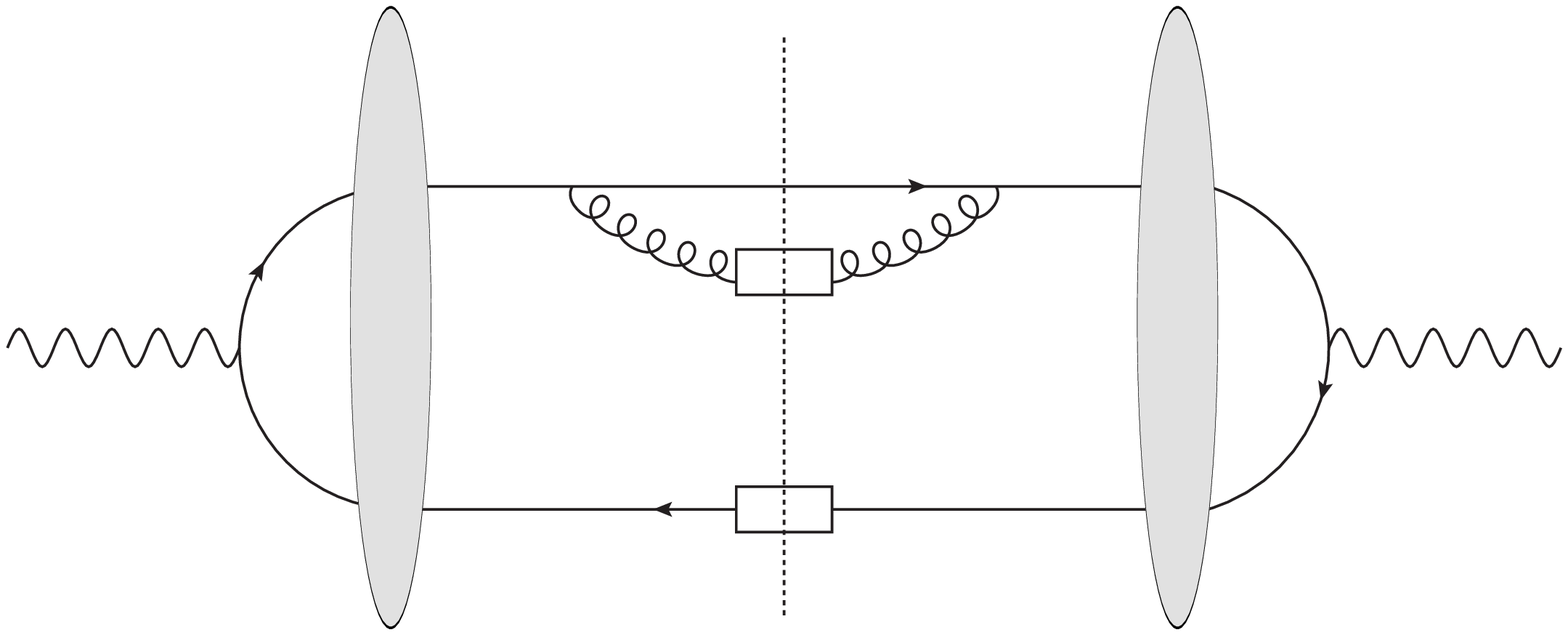}}
\put(208,180){\small (6) : collinear ($qg$)}
\end{picture}
\vspace{-185pt}
\caption{Divergent diagrams in the diagram (b), (c), and (d) of fig~ \ref{fig:NLO_FF}. Diagrams (1-4) correspond to the divergent part of diagram (b), diagram (5) is the divergent diagram in diagram (c), and (6) for (d). }
\label{fig:div_diagram}
\end{figure}

To deal with divergences, dimensional regularization, and an IR cut-off $\alpha$ are used for the transverse and longitudinal integrations respectively. Soft and collinear are the only divergences present and are contained in $d \sigma_{1 JI}$ and $d\sigma_{3JI}$. The rapidity divergences, proportional to some $\ln \alpha$ terms have been removed at the level of amplitude using the  B-JIMWLK equation, as explained in \cite{Boussarie:2016ogo}.  
Diagram (e) in fig~\ref{fig:NLO_FF} corresponds to the counterterms produced by putting the FF renormalization and evolution equation taken from \cite{Ivanov:2012iv}
%of the FF with the usual splitting function  
%\begin{equation}
%\begin{aligned}
 %    D_{q}^{h}(x) &  =D_{q}^{h}\left(x, \mu_{F}\right)-\frac{\alpha_{s}}{2 \pi}\left(\frac{1}{\hat{\epsilon}}+\ln \frac{\mu_{F}^{2}}{\mu^{2}}\right) \int_{x}^{1} \frac{d \beta}{\beta}\left[D_{q}^{h}\left(\frac{x}{\beta}, \mu_{F}\right) P_{q q}(\beta) \right. \\ 
  %   &  \left. + D_{g}^{h}\left(\frac{x}{\beta}, \mu_{F}\right) P_{gq}(\beta)\right]
   %  \end{aligned}
%\end{equation}
into the LO cross-section eq~\eqref{LO cross_section}. 
%Here, $\frac{1}{\hat{\epsilon}} = \frac{\Gamma\left(1-\epsilon\right)}{(4\pi)\epsilon} = \frac{1}{\epsilon} + \gamma_E - \ln 4\pi + \mathcal{O}(\epsilon)$.
%
Collinear divergences can only come from  diagrams where the splitting occurs after the shockwave and the same is true for soft divergences, see fig~\ref{fig:div_diagram}.

The collinear divergences appear as denominators of $\left(x_i'\vec{p}_g - x_g \vec{p}_i \right)^2$ with $i \in \{q, \bar{q}\}$ in $d \sigma_{3JI} $. 
To extract those divergences, we need to Fourier transform the non-perturbative part to disentangle and integrate over the spectator parton  (the non-fragmenting one) transverse momentum easily. We also need to change variables from $(x_i', x_g)$  to $(x_i, \beta)$ where $(x_i', x_g), x_i$ are the longitudinal fractions of the \textit{children} and \textit{parent} partons wrt to the photon momentum and $\beta$ is the longitudinal fraction wrt to the parent parton. This is to be able to compare to the counterterms. 
When extracting the divergent part of diagrams (1) and (3) of fig~\ref{fig:div_diagram}, one has to introduce the + prescription and  remove the resulting soft contribution to avoid double counting. This issue does not appear for diagrams (5) and (6). 

The soft contribution of diagram (b) of fig~\ref{fig:NLO_FF} is computed from diagrams (1-4) of fig~\ref{fig:div_diagram} altogether. We rescale  $\vec{p}_g = x_g \vec{u}$ with $|\vec{u}| \sim |\vec{p}_h| $ to isolate the divergences in the form of $\int_{\alpha}^1 \frac{d x_g}{x_g^{3-d}}$. In the rest of the integrand, we put safely  $x_g$ to 0 (as $x_q', x_{\bar{q}}'$ cannot be arbitrarily small, being limited by $x_{h}$).  
Similar changes of variables as in the collinear case are realized too. 
Most of the soft divergences in (1-4) cancel with diagram (a) of fig~\ref{fig:NLO_FF}. The rest cancel with divergences introduced by the $+$ prescription in (1) and (3). 
The leftover divergences from diagrams in fig~\ref{fig:div_diagram} cancel with the counterterms. 

\section{Conclusion and Outlook}
We have computed the NLO cross-sections of the  diffractive  production of a pair of hadrons with large $p_T$ out of $\gamma^{(*)} + p/A $ for all possible sets of photon polarization and in general kinematics $ (Q^2, t, p_T) $. Divergences have been cancelled altogether between the counterterms from the FF renormalization and evolution equation, dipole $\times $ dipole real, and virtual cross-sections. They are applicable to both the LHC with Ultra-Peripheral collisions and to the Electron-Ion Collider.

%\section*{Ackowledgements}

%\bibliographystyle{apsrev}
\bibliographystyle{JHEP}
\bibliography{ref}

\end{document}